\documentclass[aip,reprint]{revtex4-1}
\usepackage{blindtext}
\usepackage{amsmath}
\usepackage{graphicx}
\usepackage{adjustbox}
\usepackage{multirow}
\usepackage{booktabs}
\usepackage{color,soul}
\usepackage{tabularx}
\usepackage[table,xcdraw]{xcolor}
\usepackage{url}
\usepackage[colorlinks,
            linkcolor=blue,
            anchorcolor=blue,
            citecolor=blue,
            ]{hyperref}
\usepackage{setspace}
\begin{document}

%\draft % marks overfull lines with a black rule on the right

%\begin{spacing}{2.0}

\title{Microwave Applications of Photonic Topological Insulators}

\author{Shukai Ma}
\email{skma@umd.edu}
\affiliation{Quantum Materials Center, Department of Physics, University of Maryland, College Park, Maryland 20742-4111, USA}
\author{Steven M. Anlage}
\affiliation{Quantum Materials Center, Department of Physics, University of Maryland, College Park, Maryland 20742-4111, USA}
\affiliation{Department of Electrical and Computer Engineering, University of Maryland, College Park, Maryland 20742-3285, USA}

%\date{\today}

\begin{abstract}
This Perspective examines the emerging applications of photonic topological insulators (PTIs) in the microwave domain.
The introduction of topological protection of light has revolutionized the traditional perspective of wave propagation through the demonstration of backscatter-free waveguides in the presence of sharp bending and strong structural defects.
The pseudospin degree of freedom of light enables the invention of unprecedented topological photonic devices with useful functionalities.
{Our aim is to present a brief introduction of recent developments in microwave PTI demonstrations.  We give a clear comparison of different PTI realizations, summarize the key features giving rise to topological protection, and present a discussion of advantages and disadvantages of PTI technology compared to existing microwave device technology.  We conclude with forward-looking  perspectives of how the advantages of this technology can best be exploited.}
%{Here we aim to present a brief introduction of the resent development in PTI demonstrations from a more technical point of view.}
%{We next give a clear comparison among different PTI realizations of the key features giving rise to the topological protection, following with a discussion of the current PTI technology and future perspectives.}
\end{abstract}


\maketitle %\maketitle must follow title, authors, abstract and \pacs

%\section {Introduction}

\text{Introduction.---}Topological photonics, an emerging $21^{st}$ century technology, has enjoyed a boost of research interest from both the physics and engineering communities \cite{Lu2016,Khanikaev2017,Ozawa2019}. 
Similar to its electronic counterpart \cite{Hasan2010}, photonic topological insulators (PTIs) possess a full photonic bandgap which restricts light propagation in the bulk region of the structure. Topologically protected edgemodes are allowed to propagate at the boundary of the structure, creating an alternative type of waveguide structure. 
The propagation of the edgemode is free from back-scattering arising from disorder as long as it does not destroy the symmetry giving rise to the topological properties.
Inside the waveguide, the robustness of edgemode propagation is demonstrated by having waves passing around sharp corners and across disordered regions.
{In addition, the edgemodes have new degrees of freedom that can be used to ``sort'' the modes, and this can be used to make decisions and route them differently.}
Topological photonics has led to the invention of photonic devices with unprecedented functionalities, as well as upgrades to the performance of existing photonic devices. 
{A series of papers have provide a thorough review of different kinds of PTI realizations as well as the physical origin of their non-trivial topological phases} \cite{Lu2016,Khanikaev2017,Ozawa2019}.
{Here we aim to share our perspectives of the application potential of PTI technology.
We start with a brief review of current PTI demonstrations, and then discuss the emerging and future microwave applications based on those demonstrations and realistic proposals.}

Topological insulators (TIs) were discovered in the context of electronic properties of many-electron condensed matter systems.  
However it was later realized that the basic concepts underlying TI phenomena are quite general, do not depend in an essential way on collective electronic or quantum phenomena, and therefore apply to various classical wave phenomena \cite{Raghu2008}. 
A similar mapping of concepts was realized some time ago in the field of quantum chaos (or wave chaos) where ideas from nuclear physics and mesoscopic condensed matter physics crossed over to classical wave phenomena, and gave rise to many clear demonstrations of fundamental wave chaotic phenomena such as universal statistical fluctuations in microwave analogs of the Schrodinger equation \cite{Stockmann1999,Hemmady2005,Hemmady2006}.
For TI analogs one can exploit the same symmetries as those found in electronic systems to create photonic topological edge states.  
One can also create analogs of the spin-1/2 degree of freedom of the electron, or introduce artificial gauge fields that enable topologically-protected waveguide modes.
However, the photonic analogs are Bosonic in nature and lack properties arising from the Pauli exclusion principle and Fermi-Dirac statistics.

{\def\arraystretch{1.4}\tabcolsep=9pt

\begin{table*}[]
\centering
\resizebox{\textwidth}{!}{%
\begin{tabular}{@{}clll@{}}
\hline
\textbf{PTI Category}                                                             & \multicolumn{1}{c}{\textbf{Example Realizations}}                                                    & \multicolumn{1}{c}{\textbf{\begin{tabular}[c]{@{}c@{}}Mechanism to Create \\ Topological Light\end{tabular}}} & \multicolumn{1}{c}{\textbf{Binary Degrees of Freedom}}                                                  \\ \hline
\begin{tabular}[c]{@{}c@{}}Quantum Hall\\ (QH)\end{tabular}                       & \begin{tabular}[c]{@{}l@{}}Gyro-magnetic photonic crystal\\ with biasing magnetic field\cite{Wang2009,Poo2011,Ma2019} \end{tabular} & \begin{tabular}[c]{@{}l@{}}T-symmetry breaking by magneto-\\ optical effect\cite{Wang2009} \end{tabular}                      & \begin{tabular}[c]{@{}l@{}}Circulation direction controlled \\ by biasing H-field direction\end{tabular} \\ \cline{2-4}
\begin{tabular}[c]{@{}c@{}}Quantum valley\\ Hall (QVH)\end{tabular}               & BMW lattice \cite{Gao2017,Kang2018}                                                                                         & \begin{tabular}[c]{@{}l@{}}Lattice symmetry reduction by breaking \\ in-plane inversion symmetry\end{tabular} & \begin{tabular}[c]{@{}l@{}}Valley degrees of freedom consti-\\ tuted by circular polarized states\end{tabular}          \\ \cline{2-4}
\multirow{3}{*}{\begin{tabular}[c]{@{}c@{}} \\  \\ Quantum spin\\ Hall (QSH)\end{tabular}} & Ring resonator lattice \cite{Gao2016,Hafezi2013}                                                                              & \begin{tabular}[c]{@{}l@{}}Different coupling lengths between \\ the rings\end{tabular}                       & \begin{tabular}[c]{@{}l@{}}Degenerate whispering gallery-\\ like modes\end{tabular}                     \\
                                                                                  & {BMW lattice}  \cite{Cheng2016,Ma2019,Slobozhanyuk2019}                                                                                        & Bi-anisotropy effect                                                                                          & \begin{tabular}[c]{@{}l@{}}Combinations of degenerate TE\\ and TM modes\end{tabular}                     \\
                                                                                  & Hexagonal lattice  \cite{Wu2015,Yang2018}               
                                                                                  & \begin{tabular}[c]{@{}l@{}}Lattice symmetry reduction by \\ distorting the unit cell\end{tabular}             & \begin{tabular}[c]{@{}l@{}}Combination of TM modes with \\ different $E_z$ profiles\end{tabular}          \\ \\ \cline{2-4} 
Floquet-PTI                                                                       & \begin{tabular}[c]{@{}l@{}}Single mode waveguide array \\ modulated in z-direction \cite{Rechtsman2013a} \end{tabular}      & \multicolumn{2}{l}{\begin{tabular}[c]{@{}l@{}}Breaking z-reversal symmetry with helical waveguides modulated in the\\ z-direction of the x-y 2D lattice.\end{tabular}}                                                  \\ \hline

\end{tabular}
}
\caption{\label{tab:table} Summary of the general concepts of a selected subset of photonic topological insulator (PTI) experimental realizations in 2D and quasi-2D lattices. A binary degree of freedom (spin/valley) is observed in the quantum spin/valley-Hall PTIs. The abbreviation BMW stands for bi-anisotropic meta-waveguide.}
\end{table*}

}

%\section{Selected Realizations of Photonic Topological Insulator}

Selected realizations of photonic topological insulators.---The realizations considered here essentially involve a two-dimensional lattice that gives rise to one-dimensional edge states.
The quantum Hall (QH) effect is the pioneer example of topological systems in condensed matter physics.
Under low temperature and large magnetic field, a 2D electronic system with insulating bulk can acquire a non-zero topological invariant.
At the boundary formed by the interface of two insulators with different topological indices \cite{BlancodePaz2019}, the change of the topological invariant leads to the closing of the bandgap \cite{Hasan2010}, while the two bulk regions remain gapped.
Thus edge states are formed and are localized at the interfaces.
These states typically disperse in momentum, giving rise to a finite propagation velocity.
The total number of edge modes depend on the difference between the topological indices of both media.

\begin{figure*}
\centering
\includegraphics[width=1\textwidth]{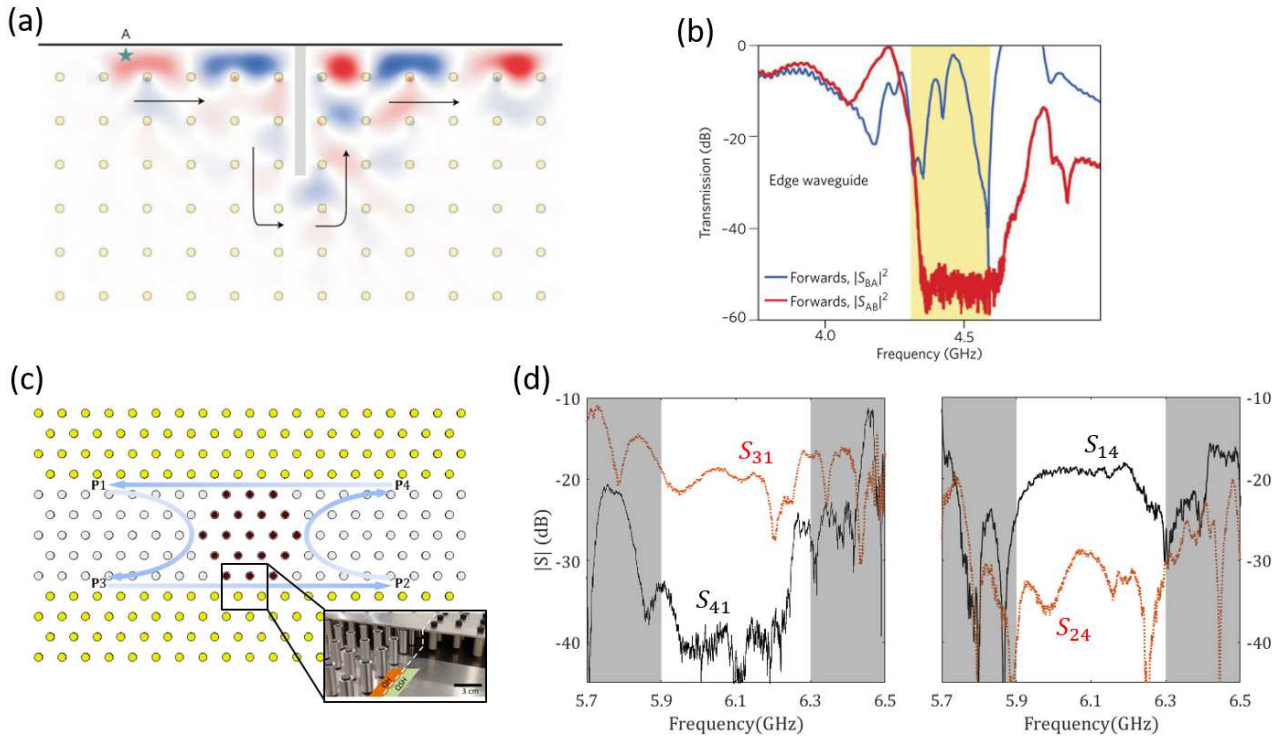}
\caption{\label{fig:fig1} {Selected experimental demonstrations of microwave PTI-based devices.}
\textbf{(a)}, {The edgemode propagation is robust against the insertion of a metallic defect (grey rectangle) at the boundary of a QH-PTI built with a lattice of biased ferrite rods.} 
The yellow circles represent the lattice sites. The blue and red colors are the out-of-plane component of the electric field. \textbf{(b)}, Transmission measurements of the edgemode propagation in \textbf{(a)}. Here $S_{BA}$ and $S_{AB}$ are a measure of the forward and backward transmission along the edge, respectively. A $\sim 0dB$ maximum forward transmission and a $\sim 50dB$ right-left difference are observed, showing the uni-directional propagation of the edgemode inside the photonic bandgap (yellow).
\textbf{(c)}, Schematic of the 4-port BMW-PTI circulator. The circulator is built with three QSH regions with alternating spin-Chern numbers (colored as yellow, gray and yellow), and a center QH region (black) biased by an external H-field pointing out of the plane. The edgemode circulates around the central QH-island from port 4$\rightarrow$1$\rightarrow$3$\rightarrow$2$\rightarrow$4, shown as blue arrows. The inset is an open-plate view of the composite QH-QSH interface.
\textbf{(d)}, Transmission measurements of the realized 4-port circulator in \textbf{(c)}. In the bulk bandgap region, the outgoing waves from ports 1 and 4 are directed to ports 3 and 1, respectively.
Panels reproduced with: \textbf{(a)},\textbf{(b)}, ref. \cite{Wang2009}, permission from Nature 461, 772–775 (2009). Copyright 2009, Macmillan Publishers Ltd;
\textbf{(c)},\textbf{(d)}, ref. \cite{Ma2019}, permission from Physical Review B 100, 085118 (2019). Copyright 2019, APS.}
\end{figure*}

We present in Table. \ref{tab:table} a brief summary of the core features of the major two-dimensional PTI genres. 
One of the pioneer PTI realizations utilized the translation of the quantum Hall (QH) effect from the electronic to the photonic domain. 
The microwave photonic crystal lattice consists of gyro-magnetic ferrite rods biased by static magnetic fields \cite{Wang2009, Poo2011}. 
The structure breaks time-reversal (T) symmetry, and in analogy with the QH effect, a chiral edgemode is observed to propagate uni-directionally at the boundary of the insulating bulk (shown in Fig. \ref{fig:fig1} (a), (b)).
The circulation direction of the edgemode is dictated by the magnetic bias field direction and the edgemode is demonstrated to flow around defects, such as a metallic impediment, with no back-scattering.
{Multimode one-way waveguides were also experimentally demonstrated by utilizing bands with large differences in Chern numbers} \cite{Skirlo2015}.
However, gyromagnetic materials are lossy and providing a DC magnetic bias is somewhat inconvenient.
It is of keen interest to search for alternative PTI realizations that do not require external magnetic field.

Another condensed matter system that shows topological edgemodes in the absence of an external magnetic field is the quantum spin-Hall (QSH) effect.
Instead of breaking the T-symmetry as in the QH effect, QSH requires the presence of T-symmetry such that Kramers degeneracy is preserved.
Electrons with opposite spins form helical edge states with opposite signs of momentum.
Naturally, finding the photonic analog of the QSH topological phase provides an alternative pathway to realize PTIs which are not restricted to gyro-magnetic materials and a biasing magnetic field. 
The QSH requires an analog of the spin-1/2 degree of freedom and a spin-orbit (or spin-momentum locking) phenomenon to be present.
The QSH-PTIs are achieved by creating an artificial spin degree of freedom (DOF) for photonic modes \cite{Khanikaev2017}.
The early proposals of T-invariant PTIs were realized with a lattice of coupled ring resonators \cite{Hafezi2013,Gao2016}. 
This design translates the two spin DOFs of an electron into the two circulation directions of the light in the ring (the clockwise or counter-clockwise modes). 
In Ref. \cite{Hafezi2013}, two circulating modes experience different optical paths when travelling from one site resonator to another.
An effective uniform magnetic field is realized if one arranges the rings into a 2D lattice, where edge modes are excited by a laser field along the structural boundary.

A number of ways have been proposed to replicate the spin-1/2-like degree of freedom in the photonic context.
Here we concentrate on a selected few of these realizations.
QSH-PTIs can be realized with the so-called bianisotropic meta-waveguide (BMW) systems \cite{Ma2015,Slobozhanyuk2019,Cheng2016,Ma2017}. 
One starts with a perfect 2D hexagonal metal-post photonic crystal sandwiched between two conducting plates with carefully engineered degeneracy of the TE and TM modes at the Dirac points ($K$ and $K'$ points) in the 2D photonic band structure.
The spin-1/2 degree of freedom is created by two orthogonal linear combinations of the TE and TM modes at the Dirac points. 
By breaking the inversion symmetry in the perpendicular ($z$) direction of the lattice, a bianisotropic coupling emerges between the transverse electric (TE) and transverse magnetic (TM) modes, which corresponds to an extra magneto-electric mixing term between electric and magnetic fields ($\textbf{D}=\hat{\epsilon} \textbf{E}+\hat{\chi}\textbf{H}$ and $\textbf{B}=\hat{\mu} \textbf{H}+\hat{\chi}^{\dagger}\textbf{E}$, where $\hat{\chi}$ is the magneto-electrical coupling parameter).
The non-vanishing off-diagonal terms of $\hat{\chi}$ play a similar role to the off-diagonal components of a gyro-magnetic material's permeability tensor, which is responsible for emulating an artificial gauge field in QH-PTIs.
In BMW systems, the bianisotropic effect introduces an artificial gauge field and Berry connection for the two spin modes and further gives rise to a QSH-PTI \cite{Ma2015,Xiao2016}. 
{With the application of QSH-PTIs, robust one-way edgemode propagation can be realized without the application of external magnetic field} \cite{Lu2016}. 
{This effectively reproduces the functionality of an isolator, but without the need for magneto-optical materials or the concern of a finite VSWR inherent in all conventional microwave devices.  One important benefit of QSH-PTIs is in applications that suffer from environmental magnetic fields, such as in superconducting quantum computing devices.}

It was later shown that the BMW system is also able to host QH \cite{Ma2019} and quantum valley-Hall (QVH) \cite{Gao2017,Kang2018} topological phases through the inclusion of magneto-optical components, or an in-plane inversion-breaking tripod structure, respectively. 
An outstanding benefit of BMW-PTIs arises from the fact that they are all perturbations of the same underlying photonic crystal structure.
This allows for the demonstration of composite PTI systems, where different topological phases co-exist to create unique edgemodes, and thus perform useful functionalities, such as a unique 3-port Y-junction and a full 4-port circulator \cite{Ma2019} (shown in Fig. \ref{fig:fig1} (c), (d)).

\begin{figure*}
\centering
\includegraphics[width=0.85\textwidth]{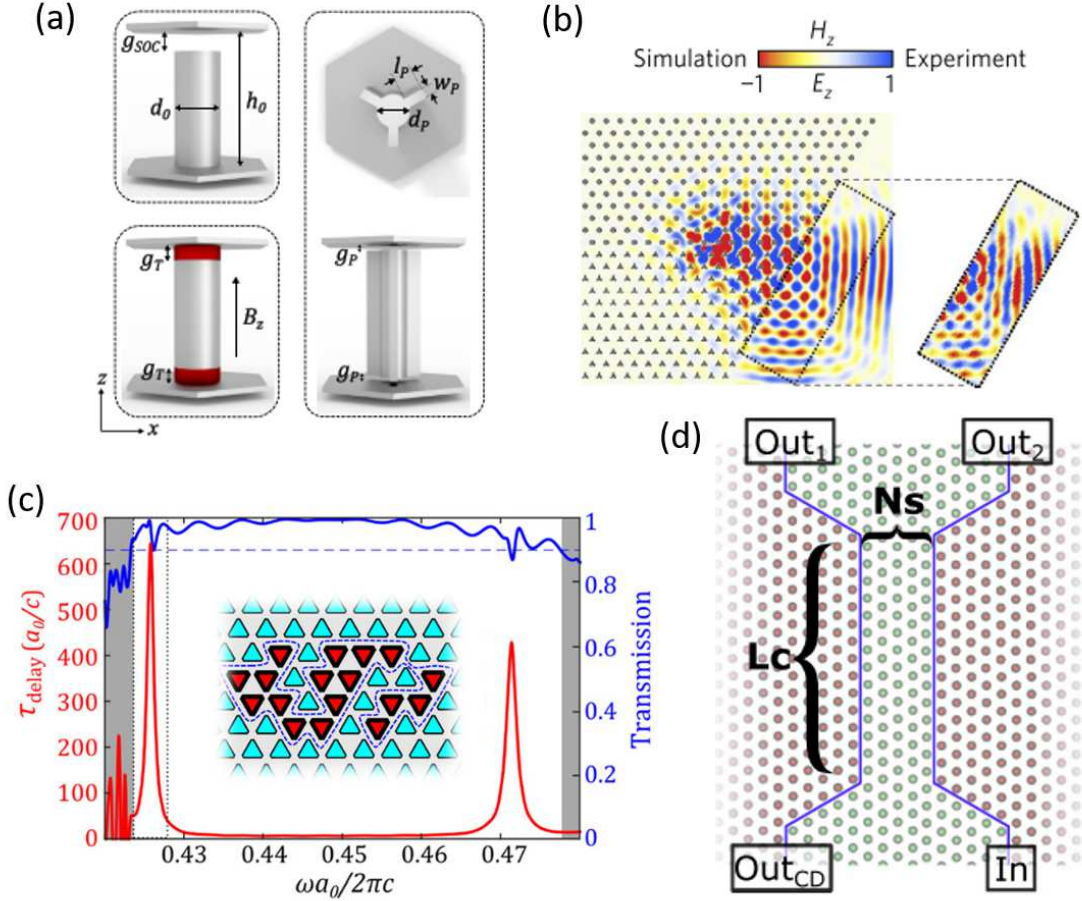}
\caption{\label{fig:bmw} {BMW-PTIs and selected microwave applications.}
\textbf{(a)}, Unitcell schematics of QSH, QH and QVH-PTIs based on the BMW system, respectively. Here $d$, $h_0$ and $g$ are the diameter, unitcell height and spacing distance. $B_z$ denotes the z-component of the magnetic field used to bias the ferrite disks (red) in a QH-BMW structure, while $l_P$ and $w_P$ denote the structure details of the QVH tri-pods. 
\textbf{(b)}, The field profile of the leaky wave radiated from a QVH-BMW waveguide termination in both simulation and experiment. \textbf{(c)}, Transmission and delay-time simulation results of the delay-line structure which is realized by a random QVH-BMW cavity. The inset shows a schematic of the random cavity. \textbf{(d)}, Schematic of a directional coupler based on two parallel coupled QSH-BMW waveguides. Quantities $L_c$ and $N_s$ refer to the interaction length and the inter-waveguide separation. Red and green rods represent different QSH regions. 
Panels reproduced with: \textbf{(a)}, ref. \cite{Ma2017}, permission from Physical Review B 95, 165102 (2017). Copyright 2017, APS; \textbf{(b)}, ref. \cite{Gao2017}, permission from Nature  Physics 14, 140–144 (2018). Copyright 2018, Macmillan Publishers Ltd; \textbf{(c)}, ref. \cite{Ma2016}, permission from Tzuhsuan Ma and Gennady Shvets, New Journal of Physics 18, 025012 (2016). Licensed under a Creative Commons Attribution (CC BY 3.0) license; \textbf{(d)}, ref. \cite{Gentili2019}, permission from Physical Review B 100, 125108 (2019). Copyright 2019, APS.}
\end{figure*}

QSH-PTIs are also realized in distorted hexagonal crystalline lattices of scatterers \cite{Wu2015,Yves2017}. The combination of TM modes with $p/d-$like symmetry $E_z$ profiles constitutes the two pseudospins, and the expansion of the hexagonal unit cell leads to band inversion and non-trivial topological bands.
The contraction of the same unit cell creates a topologically trivial insulating bulk.
Localized edge states are thus created at the interface of the two regions.

An alternative approach to creating topologically protected modes is the so-called Floquet TIs.
Floquet systems involve periodically-driven structures, either in space or in time.
Consider first spatially-modulated systems in a quasi-2D lattice \cite{Rechtsman2013a}.
The system consists of a honeycomb lattice of coupled waveguides which extend in a helical manner in the perpendicular$(z)$-direction to the 2D lattice plane. 
This spatial modulation of the waveguides, effectively equivalent to applying a fast and strong temporal modulation of a strictly 2D system \cite{Fang2012}, generates a synthetic gauge field and further gives rise to the opening of a photonic bandgap.
Temporal modulation has enabled various exciting photonic applications \cite{Estep2014,Hu2015}.
However the experimental demonstration of Floquet topological insulator in time awaits investigation due to the challenge of implementing sufficiently strong modulation in practice \cite{Khanikaev2017}.

{\def\arraystretch{1.4}\tabcolsep=2.5pt

\begin{table*}[]
\centering
\resizebox{\textwidth}{!}{%
\begin{tabular}{ccccllcl}
\hline
\textbf{PTI Platform}                                                         & \textbf{$f_{op}$} & \textbf{FBW} & \textbf{\begin{tabular}[c]{@{}c@{}}$a_0/\lambda_{op}$\\ mm/mm\end{tabular}} & \multicolumn{1}{c}{\textbf{\begin{tabular}[c]{@{}c@{}}Potential\\ Destructive Disorder\end{tabular}}}                  & \multicolumn{1}{c}{\textbf{Remarks}}                                                                    & \textbf{\begin{tabular}[c]{@{}c@{}}Compatibility \\ with Other PTIs\end{tabular}} & \multicolumn{1}{c}{\textbf{Potential Application}}                                                               \\ \hline
\begin{tabular}[c]{@{}c@{}}Gyro-magnetic \\ QH-PTI\cite{Wang2009} \end{tabular}               & 4.5GHz            & 6\%          & 40/66                                                                       & \multicolumn{1}{c}{Magnetic disorder}                                                                                      & \begin{tabular}[c]{@{}l@{}}Magneto-optical effects \\ are hard to emulate at \\ higher frequencies\end{tabular} & No                                                                                & \begin{tabular}[c]{@{}l@{}}Isolators, \\ slow-light devices, \\ bandpass filters ...\end{tabular}                \\
\begin{tabular}[c]{@{}c@{}}Coupled ring \\ resonators \\ QSH-PTI\cite{Gao2016} \end{tabular} & 11.3GHz           & 5\%          & 83/28                                                                       & \begin{tabular}[c]{@{}l@{}}Spin-flipping defects; \\ Strong dissipative \\ disorder\end{tabular}                      & \begin{tabular}[c]{@{}l@{}}Fine fabrication \\ techniques are required \\ to prevent losses\end{tabular}           & No                                                                                & \begin{tabular}[c]{@{}l@{}}Topological mode \\ amplification ...\end{tabular}                                    \\
\begin{tabular}[c]{@{}c@{}}Distorted lattice\\ QSH-PTI\cite{Yang2018} \end{tabular}         & 7.5GHz            & 2\%          & 25/40                                                                       & \begin{tabular}[c]{@{}l@{}}Sensitive to the ratio \\ between lattice constant \\ and contraction distance\end{tabular} & \begin{tabular}[c]{@{}l@{}}Band gap opened at $\Gamma$ \\ point, allowing flexible \\ propagation directions  \end{tabular}                                    & No                                                                                & \begin{tabular}[c]{@{}l@{}}Applicable to both \\ metallic and dielectric \\ realizations\end{tabular}            \\
\begin{tabular}[c]{@{}c@{}}BMW Composite\\ PTI systems\cite{Gao2017,Ma2019} \end{tabular}           & 6GHz              & 6\%          & 36/50                                                                       & \begin{tabular}[c]{@{}l@{}}Spin-flipping or inter-\\ valley scattering disorder\end{tabular}                          & \begin{tabular}[c]{@{}l@{}}Modes propagate along \\ $K/K'$ directions\end{tabular}                                  & Yes                                                                               & \begin{tabular}[c]{@{}l@{}}Circulators, isolators, \\ photonic logic devices, \\ high-power devices\end{tabular} \\ \hline
\end{tabular}%
}
\caption{\label{tab:table2} Practical metrics for a selected subset of PTI microwave realizations. The column $f_{op}$ includes the reported central operating frequency of example PTI realizations. FBW stands for the fractional bandwidth, defined as $FBW=\Delta f/f_{op}$ where $\Delta f$ is the size of the bandgap. The quantities $\lambda_{op}$ and $a_0$ represent the operating wavelength and the lattice constant of the photonic crystal.}
\end{table*}

}
%\section{Properties of Selected PTI Realizations}

\text{Properties of selected PTI realizations.---}We briefly introduced three main PTI approaches: time-reversal breaking (QH), time-reversal invariant (QSH, QVH) and the Floquet topological systems. 
Through careful design, each TI system possess insulating bulk regions and edge/kink states at the structure boundary.
The properties of the edge state, such as the momentum, spin index and the total number of states, are based on the topological order of the bulk.
Unique transmission properties are promised in various PTIs. 
This includes backscatter-free and uni-directional propagation involving highly localized edge states. 
These properties are maintained as long as the underlying symmetry is not violated.

A brief summary of practical specifications of selected microwave topological phases can be found in Table \ref{tab:table2}.
For the four exemplary PTI realizations, the topologically protected modes all operate in the GHz range with a relatively narrow fractional bandwidth (FBW).
The physical dimension of the system as compared to the operating wavelength is another important concern for practical uses.
In the coupled ring resonator system, the diameter of the ring resonators is large compared to the wavelength of the light \cite{Hafezi2013,Gao2016}. 
On the other hand, the crystalline and BMW based PTIs are more compact in size. 
Apart from the microwave realizations, the concept of topological protection of modes can also be accommodated in a wide variety of physical wave systems. 
Examples range from acoustic systems \cite{Khanikaev2015}, electrical circuits \cite{Zhu2019}, mechanical systems \cite{Huber2016} and optical lattices \cite{Harari2018}.

%\section{An Example System: Bi-anisotropic Meta-Waveguide Photonic Topological Insulators}

{An example system: Bi-anisotropic Meta-Waveguide PTIs.---}The BMW-PTIs are based on a photonic crystal formed from an unperturbed unit cell consisting of metallic cylindrical rods sandwiched between two metallic plates \cite{Ma2015,Xiao2016}. 
{A purely dielectric version has also been demonstrated} \cite{Slobozhanyuk2019}.
The unit cells are arranged into a hexagonal 2D lattice. 
The dimensions of the structure are carefully engineered so that the TE and TM modes are degenerate at the Dirac points in the photonic crystal band structure \cite{Ma2015}.
A variety of perturbations are able to break this mode degeneracy to create a bandgap in the bulk, and to give rise to different topological states for the material. 
The synthetic spin DOFs are formed by the in-phase and out-of-phase combination of the TE and TM modes \cite{Ma2015}. 
The propagation of these spin-labelled modes is along the $K$ and $K'$ directions of the lattice.
With the introduction of inversion-symmetry-breaking in the vertical ($z$) direction of the lattice, an effective spin-orbit coupling (SOC) interaction is created and non-trivial topological indices are assigned to the two spin modes.
The QSH analog photonic topological systems are thus realized without the application of a global magnetic field.
The propagation of edgemodes in QSH-BMWs are experimentally observed at the interface between PTIs with different topological indices in various experiments \cite{Cheng2016,Xiao2016}. Like its electronic counterpart, the momentum of a photonic edgemode is locked to its synthetic spin polarization. The edgemode propagation is free from back-scattering as long as the underlying T-symmetry of the system is preserved and severe loss is absent.

Recent studies reveal the possibilities of emulating both QH and QVH topological phases using the BMW architecture \cite{Ma2017,Xiao2016,Kang2018,Gao2017}. The QVH-BMWs are realized by substituting the cylindrical center rod into a carefully designed tri-pod which creates the in-plane parity symmetry breaking \cite{Gao2017,Kang2018} (e.g., reflect the tri-pod with respect to the $x$-direction).
The valley refers to the vicinity of two high symmetry points $K/K'$. 
The QH-BMWs are introduced with the application of magneto-optical materials biased by magnetic fields \cite{Ma2019} which breaks the T-symmetry and introduces a bandgap.
The topological index of the QSH, QVH and QH BMW modes are defined with the spin-Chern number $2C_{s,v}^{SOC} = \pm sgn(\Delta_{SOC})$, valley-Chern number $2C_{s,v}^{P} = \pm sgn(\Delta_{P})$, and a global Chern number $2C_{s,v}^{T} = sgn(\Delta_{T})$, respectively \cite{Ma2017}. 
Here $s=\uparrow / \downarrow$ is the spin label and $v=K/K'$ is the valley label in the above expressions. The $\Delta_{SOC}$, $\Delta_P$ and $\Delta_T$ are the overlap integrals of the unperturbed modes inside the perturbed volume of the unit cell, whose values set the scale for the width of the bulk bandgap. The three subscripts of $\Delta$ represent the specific types of symmetry-breaking which lift the mode degeneracy at the Dirac points and further give rise to non-trivial topological phases. 

As mentioned before, heterogeneous PTI structures have recently been demonstrated with QSH, QVH and QH BMW lattices as building blocks \cite{Gao2017,Ma2019}. 
A key requirement for a successful composite topological structure is the matching of the above $\Delta$ perturbation integrals. 
This ensures the reflection-free propagation of edgemodes when travelling through a heterogeneous interface between different PTI phases. 
Inside a composite topological device, the propagation properties of an edgemode is dictated by its spin index \cite{Ma2019}: the number of edgemodes are determined by the difference in Chern numbers between two neighboring media, and the propagating direction of the edgemode is defined by both its spin index and the polarization direction of the QH-BMW region.
Practical photonic devices, such as a 4-port circulator, have been experimentally realized with the seamless combination of QH and QSH BMW topological phases \cite{Ma2019} (Fig. \ref{fig:fig1} (c) and (d)). 
The structure consists of four I/O channels made by QSH-QSH waveguides and a center QH island to shuttle and dispatch the flow of edgemodes. 
High isolation is promised by the topologically protected propagation of edgemodes.
Real-time switching of circulation direction is also achievable with the simple inversion of the biasing H-field.
{A 3-port junction has also been experimentally realized utilizing a QH-QSH composite structure. 
The S-parameters of this passive device directly mimic those of a quasi-circulator, which ordinarily require active components} \cite{Ma2019}.
Apart from being a demonstration of a PTI-based practical application, this device is also the experimental realization of a composite physical material with both QSH and QH topological phases, which is unprecedented in either an electronic or photonic setting. 
Recently a BMW-based compact and scalable delay line structure was experimentally demonstrated \cite{Lai2016}.
{All-dielectric bi-anisotropic structures provide alternative approaches to realize low-loss on-chip BMW-based PTI systems} \cite{Slobozhanyuk2019}.

{\def\arraystretch{1.5}\tabcolsep=8pt

\begin{table}[]
\centering
\resizebox{0.5\textwidth}{!}{%
\begin{tabular}{ccc}
\hline
\textbf{\begin{tabular}[c]{@{}c@{}}Exemplary\\ PTI Device\end{tabular}} & \textbf{\begin{tabular}[c]{@{}c@{}}Specifications\\ (PTI/Conventional)\end{tabular}} & \textbf{Characteristics}                                                                     \\ \hline
Isolator\cite{Wang2009}                                                                & \begin{tabular}[c]{@{}c@{}}Isolation: \\ $\sim50/18$dB\end{tabular}                  & High isolation                                                                               \\
Circulator\cite{Ma2019}                                                              & \begin{tabular}[c]{@{}c@{}}Isolation: \\ $\sim25/18$dB\end{tabular}                  & \begin{tabular}[c]{@{}c@{}}Real-time circulation \\ direction switching\end{tabular}         \\
Antenna\cite{Gao2017}                                                                 & \begin{tabular}[c]{@{}c@{}}Immunity to \\ impedance mismatch\end{tabular}       & \begin{tabular}[c]{@{}c@{}}High directivity, \\ multiple radiation\\ directions\end{tabular}                           \\
\begin{tabular}[c]{@{}c@{}}Delay-line\\ structures\cite{Ma2016} \end{tabular}         & Variable delay time                                                             & \begin{tabular}[c]{@{}c@{}}Compact \\ device size\end{tabular}  \\
\begin{tabular}[c]{@{}c@{}}Directional\\ coupler\cite{Gentili2019} \end{tabular}           & Variable coupling                                                               & \begin{tabular}[c]{@{}c@{}}Bi-directional\\ coupling\end{tabular}                                                 \\
\begin{tabular}[c]{@{}c@{}}Power\\ splitter\cite{Skirlo2014} \end{tabular}           & Tunable splitting                                                               & \begin{tabular}[c]{@{}c@{}}Unity power\\ efficiency \end{tabular}\\ \hline
\end{tabular}
}
\caption{\label{tab:table3} {Summary of exemplary PTI microwave applications that have been demonstrated experimentally or numerically. Specifications of the corresponding conventional products are adopted from off-the-shelf microwave products with similar operating bandwidth as their PTI counterparts.} }
\end{table}

}

%\section{Applications}

%\text{Applications.---}From its inception, the field of topological photonics has drawn intense research interest. 
%As compared to studying topological physics in electronic systems, photonic systems offer great convenience in designing and fabricating structures, and conducting measurements. 
%It also allows researchers to have better control of specific physical details, such as the precise introduction of another topological phases or local defects in the lattice.
%{Recent studies report the demonstration of composite photonic systems between QSH and QH/QVH phases} \cite{Gao2017,Ma2019}. 
%{Such composite topological systems are otherwise extremely hard to fabricate with electronic materials.
%Another example is the realization of topological orders in photonic amorphous systems} \cite{Zhou2020}.  
%{While in electronic systems, QSH quasi-crystals was proposed theoretically using Penrose tilt structure} \cite{Huang2018}, {whose experimental realization awaits further investigation.}
%{Moreover, the design of photonic topological systems can be facilitated by electromagnetic numerical solvers as well as top-down optimization methods} \cite{Molesky2018,Christiansen2019}.

\text{Applications.---}The attractive properties of a topologically protected mode, such as the immunity to structural defects and being able to travel through sharp corners, are of keen interest to the field of microwave engineering \cite{Khanikaev2017}. 
%In PTI-based photonic devices, the propagation of edgemodes are restricted to the boundaries of insulating bulk material \cite{Cheng2016}. 
The introduction of local disorder along a PTI interface would only alter the shape of the original bulk boundary and leave the edgemode unperturbed \cite{Wang2009}. 
Thus a PTI-based photonic device has high tolerance to structural defects, whereas the performance of more traditional microwave devices is largely limited by the fabrication techniques and the changes in the structure over time and usage.
{A clear demonstration of the superiority of PTI-based devices would be the creation of mechanically flexible waveguides.
It is commonly known that sharp bending of semi-rigid and flexible coaxial transmission lines should be strictly avoided.
However, the topologically protected modes are less susceptible to back-scattering induced from structural defects. 
This property has become a benchmark test for various PTI waveguide implementations, where the transmission of edgemode is insensitive to sharp bending of PTI waveguides} \cite{Khanikaev2012,Hafezi2013}.

The above mentioned characteristics make PTIs an excellent platform for realizing unique real-life devices.
Several microwave PTI applications have emerged and demonstrated with numerical or experimental methods (shown in Tab. \ref{tab:table3}).
{Aside from the BMW isolators and circulators (introduced in previous paragraphs)} \cite{Ma2019}, {we next focus on other exemplary PTI devices in }Tab. \ref{tab:table3} {, namely the compact delay-line structures and leaky-wave antennas.}
%{For example, the BMW based directional coupler with tunable coupling strength is recently realized numerically} \cite{Gentili2019} (Fig. \ref{fig:bmw} (d)).

BMW-based PTIs have been proposed for compact microwave delay lines \cite{Lai2016, Ma2017}.
{Such devices can be concatenated without standing wave complications (high VSWR) arising from slight impedance mismatches} \cite{Hafezi2011}.
The central operating frequency $f_{op}$ of these devices can be easily scaled by scaling the physical dimensions of the unit cell.
%{Characters of the above exemplary PTI-based microwave devices are listed in} Tab. \ref{tab:table3}.

{Unique antenna devices can be designed based on PTI devices with high directivity without impedance matching issue}.
{Robust wireless communication can be achieved by utilizing properties like the distinct refractive properties of TE and TM kink states radiated from a BMW waveguide to free space} \cite{Gao2017} (Fig. \ref{fig:bmw} (b)), {and the tunable radiation direction enabled by frequency tuning of a magnetized plasmonic structure} \cite{AliHassaniGangaraj2018}.
{Based on reciprocity, a high-performance leaky-wave emitting structure may also be a good receiver.} 
{The utilization of PTI-based wireless TX/RX system may benefit MIMO applications.}

{A common advantage of PTI devices is the immunity to impedance mismatch.
For example, a QSH-BMW waveguide may achieve near unity VSWR at both the injection and radiation ends.
At the input end,} Ref. \cite{Gentili2019} {shows that nearly ideal transmission can be achieved between the excitation source to the BMW waveguide, where $\sim1.1$ VSWR was observed across most of the operating bandwidth. 
At the output end, both numerical and experimental studies have been conducted to examine the radiation pattern of a BMW waveguide} \cite{Ma2016,Gao2017}. 
{Ideal radiation from the waveguide to an open environment is observed under the proper choice of zig-zag waveguide termination.
%Thus it is clear that the PTI applications are good for constructing devices which are insensitive to the impedance matching problem.
Note that the above examples are both conducted with BMW-PTIs, so that future studies of impedance matching properties should be done with other PTI systems. }
{In the meantime, we also note that common PTI-device problems such as the limited operating bandwidth, structural dimensions which scale with wavelength, and high insertion loss demand further optimization.}

{We next discuss the possible PTI microwave applications which await further investigation.}
%A microwave diplexer may be achieved with the QSH-QH-QSH composite BMW lattice where the external H-field applied at QH region serves as the control.} 
Metallic BMW-PTI structures based on photonic crystals are also very suitable for high-power applications because the field levels in the hexagonal array of rods are low compared to integrated structures. 
Such PTIs can be used for narrow-band high-power microwave energy delivery where impedance mismatches create strong reflected waves that can create havoc.

{An exciting possibility is to perform amplification of signals in a microwave PTI system.
Substantial efforts have made towards lasing under topological protection in the optical regime utilizing the gain-loss distributed systems with parity-time (PT) symmetry} \cite{St-Jean2017,Bahari2017,Zhao2018,Parto2018,Mittal2018}. 
Compared to topologically trivial systems, high power single mode operation is promised in a PTI based device, with high robustness to structural defects.
{PT-symmetric topological systems have been realized in the microwave regime} \cite{Poli2015}. 
{With proper inclusion of gain, such an approach could be applied to PTI-based microwave amplifiers and masers, and to provide potential integration with other PTI-enabled functionality.}
{A microwave limiter has been demonstrated using this system } \cite{Kuhl2017}.
{A topological nonreciprocal quantum-limit amplifier fabricated from a PTI structure has been proposed with parametric driving of the boundary modes} \cite{Peano2016}.

Aside from the good transmission properties, the synthetic spin of the edgemodes provides an extra handle to manipulate the flow of light. 
With the exception of polarization, this wave-steering ability is essentially unprecedented in more conventional devices. Devices with sophisticated functions can be realized by engineering the pathways of different spin DOF in a heterogeneous PTI system \cite{Ma2019}. Spin-splitters, combiners and ``logic'' devices such as Boolean networks may be realized with composite topological systems.
The spin-1/2-like DOF of PTI edge modes is like an effective 2-state bit. Inside a photonic logic device, information can be shuttled around and switched using QH-based composite PTI structures.

We also note several drawbacks of the current PTI realizations. The gyro-magnetic effect, which is responsible for breaking the T-symmetry in QH-PTI systems, typically require large magnets, and the gyromagnetic effect weakens at higher frequencies. 
Though the propagation of edgemodes are free from back-scattering induced from a variety of defects, the loss of the preserved system symmetry would cause reflection. A recent report studies a system where a metallic mirror is installed at the termination of a QVH-QVH waveguide \cite{Li2019}. 
Inter-valley scattering of edgemodes are observed which further leads to the finite reflection of the edgemode. 
The propagation of a topologically protected mode would also be destroyed under extreme dissipative defects \cite{Gao2016}.
{It is also observed experimentally that spin flipping of topologically protected modes can take place even in the presence of TR-preserving disorder} \cite{Kang2020}.
{However, the authors also show that the utilization of nonreciprocal PTI waveguide, such as gyromagnetic photonic crystal, can eliminate this disorder-induced spin flipping.}

Another inherent disadvantage of current PTI designs is that only a small portion of the structure is utilized, both in real-space and k-space.
{This limitation is not widely discussed but of utmost importance if PTIs are to be applied in real life devices.}
In real-space, there must be an insulating bulk region to provide a substrate for the `edges'.
Only a small fraction of the structure participates in the wave propagation.
{A drawback of the PTI devices is their sizes as compared to existing commercial products in similar operating frequency ranges.
The decrease in the size of the bulk region would address this limitation.}
The edgemode lateral confinement length can be estimated as $\xi \sim v_D/\Delta$, where $v_D$ is the wave velocity at the Dirac point and $\Delta$ is the direct gap \cite{Stuhler2020}.
Future work may focus on minimizing the confinement length of edgemodes into the bulk, for example by increasing the gap $\Delta$, allowing for the elimination of unnecessary bulk structures.

It is notable that enhanced energy confinement and transport can be realized with the QSH effect in planar structures, where the surface states commonly found in two-dimensional lattices are now one-dimensional line states \cite{Bisharat2019}.
As shown in the photonic band structure studies and transmission measurements, the operating bandwidth of PTI systems are small compared to the width of the conducting bands (Tab. \ref{tab:table2}).
This phenomenon is dictated by the literal analogy to the electronic topological systems.
Future generalizations of PTI wave behavior may eliminate this restriction. 
We note that a PTI delay line based on all-dielectric QVH random cavities presents a way to utilize the bulk regions to accomplish a practical goal. 
Compact delay lines are proposed utilizing the whispering-galley-like modes circulating inside the randomized bulk rather than the perimeter of the resonator \cite{Ma2016}.

%\section{Conclusion and perspectives}

\text{Conclusion and perspectives.---}We would like to conclude the paper with an outlook on topological photonics. 
A fundamental research interest of the field is to identify and exploit PTI realizations in different physical systems. 
Up to this point, many realizations of PTIs have been done in close analogy to their electronic forebearers.
Hence PTI realizations based on topological states created by more general conditions and in other dimensions are also of interest. 
Recent progress includes finding higher-order topological states \cite{Ni2018,Peterson2018}, and the demonstration of 3D PTIs either through the addition of an artificial dimension \cite{Lin2016} or in real space \cite{Jia2019,Xie2019}. 
Beyond the common photonic lattice system, topics like the experimental demonstration of topological phases in disordered systems \cite{Bandres2016,Yang2019a} await further investigation. 
{The immunity to short-range deformations has been observed in various PTI systems.}
{Robust topological transport is also validated experimentally in amorphous PTI systems with long-range deformations} \cite{Xu2020}.
The emulation of non-Hermitian PTIs with parity-time symmetric structures opens up the invention of topological systems with co-existing gain and loss, which is difficult to realize in condensed matter systems \cite{Weimann2017}.
The introduction of nonlinear effects into topological photonics would invoke unprecedented research directions in both the microwave and optical ranges, such as studying the edgemode response to weak and strong diode-induced nonlinearity \cite{Dobrykh2018}.

The application of PTI technologies are also becoming increasingly coupled with other physical research. 
{The emulation of two-state properties such as photonic spin creates another opportunity to explore spin-related physical phenomena, for example the simulation of complex scattering systems in the Gaussian Symplectic Ensemble universality class of random matrix theory} \cite{Rehemanjiang2016}.
Its superior light transmission properties make PTI technology an ideal choice in various physical experiments. 
Potential directions include emulation of fractional quantum Hall systems \cite{Umucallar2012}, the creation of photon-phonon interactions \cite{Peano2015}, and better control of the flow of light through the effective Lorentz force for photons \cite{Fang2012}.

Compared to the fundamental studies, the utilization of topologically protected light in more applied fields are less well-developed, unfortunately.
The recent advancement of heterogeneous PTI systems demonstrates the priority of realistic photonic devices that are purely empowered by topological physics \cite{Ma2019}.
As a future direction, miniature on-chip PTI devices would find broad application in fields including telecommunications and high performance all-optical computation. Though many PTI designs would be of fundamental interest, we are certain that the unique properties of topological photonic technologies would ensure its role in future applications.

%\begin{acknowledgments}
We thank B. Xiao, K. Lai and Y. Yu for photonic band structure calculations and G. Shvets for inspiring discussions.
This work was supported by ONR under Grant No. N000141912481, AFOSR COE Grant FA9550-15-1-0171, and the Maryland Quantum Materials Center.
Data sharing is not applicable to this article as no new data were created or analyzed in this study.
%\end{acknowledgments}

%\end{spacing}

%\bibliography{cit}

\end{document}